\documentclass[manuscript]{aastex}
\usepackage{amssymb}
\usepackage{txfonts}

\ifx\urlurl \undefined
\def\urlurl#1{\href{http://#1}{\textsf{#1}}}\fi
\usepackage{orcidlink}

\usepackage{amsmath}
\usepackage{float}

\shortauthors{S. Liu et al.}

\begin{document}

\title{The Magnetic Field Calibration of the {\it Full-Disk Magnetograph} onboard the {\it Advanced Space based Solar Observatory} (ASO-S/FMG)}
\author{S. Liu\altaffilmark{1,2,3}, J.T. Su\altaffilmark{1,2,3}, X.Y. Bai\altaffilmark{1,2,3}, Y.Y. Deng\altaffilmark{1,2,3}, J. Chen\altaffilmark{1,2,3}, Y.L. Song\altaffilmark{1,2,3}, X.F. Wang\altaffilmark{1,2,3}, H.Q. Xu\altaffilmark{1,2,3}, X. Yang\altaffilmark{1,2,3}  }

\affil{$^{1}$National Astronomical bservatories, Chinese Academy of Science, Beijing, 100101, China}
\affil{$^{2}$Key Laboratory of Solar Activity and Space Weather, National Space Science Center, Chinese Academy of Science, Beijing, 100190, China}
\affil{$^{3}$School of Astronomy and Space Sciences, University of Chinese Academy of Sciences, Beijing, 100101, China}

\email{lius@nao.cas.cn}

\begin{abstract}
The Full-disk magnetograph is a main scientific payload onboard the \textit{Advanced Space based Solar Observatory} (ASO-S/FMG) that through Stokes parameters' observation to measures the vector magnetic field. 
The accuracy of magnetic-field values is an important aspect of checking the quality of the FMG magnetic-field measurement. 
According to the design of the FMG, the linear calibration method under the weak-field approximation is the preferred scheme for magnetic-field calibration.
However, the spacecraft orbital velocity can affect the position of observed spectral lines, then result in a change of the polarization-signal strength. 
Thus, the magnetic field is modulated by the orbit velocity of the spacecraft. In this article, through cross calibration between 
FMG and HMI (\textit{Helioseismic and Magnetic Imager} onboard the \textit{Solar Dynamic Observatory}), the effects of spacecraft orbital velocity on the coefficient of magnetic-field calibration are investigated.
By comparing the magnetic field of FMG and HMI
with spacecraft orbital velocity as an auxiliary reference, the revised linear-calibration coefficients that depend on spacecraft orbital velocity are obtained.
Magnetic field of FMG corrected by the revised calibration coefficients removing the effect of spacecraft orbital velocity will be more accurate and suitable for scientific research.

\end{abstract}

\section{Introduction}\label{S-Introduction} 
The solar magnetic field provides an essential force that dominates solar activity. The energy and mass released from solar activity, such as solar flares, filament eruptions, corona mass eruptions and small-scale activity mostly originate from stored magnetic energy (\citeauthor{1982SoPh...79...59K}~\citeyear{1982SoPh...79...59K}; \citeauthor{1995ApJ...451L..83S}~\citeyear{1995ApJ...451L..83S}; \citeauthor{1994ApJ...424..436W}~\citeyear{1994ApJ...424..436W};
\citeauthor{1996ApJ...456..840T}~\citeyear{1996ApJ...456..840T}; \citeauthor{2000JGR...10523153F}~\citeyear{2000JGR...10523153F}; \citeauthor{2001ApJ...560..919B}~\citeyear{2001ApJ...560..919B}; \citeauthor{2002A&ARv..10..313P}~\citeyear{2002A&ARv..10..313P}; \citeauthor{2012ApJ...748L...6N}~\citeyear{2012ApJ...748L...6N}). 
The full-disk magnetograph onboard the \textit{Advanced Space based Solar Observatory} (ASO-S/FMG: \citeauthor{2019RAA....19..156G}, \citeyear{2019RAA....19..156G}; \citeauthor{2019RAA....19..157D} \citeyear{2019RAA....19..157D}) measures magnetic fields on the solar photosphere, 
which would contribute to studies of solar activity (\citeauthor{2023SoPh..298...68G}, \citeyear{2023SoPh..298...68G}).
Fe {\sc i} $\lambda$ 5234.19~\AA~is chosen as a magnetic-sensitive line by FMG to measure the photosphere magnetic field. FMG use a liquid-crystal variable retarder (LCVR) to observe polarized Stokes images ($I$, $Q$, $U$, $V$), which are calibrated to yield the magnetic field on the photosphere (\citeauthor{2019RAA....19..157D}~\citeyear{2019RAA....19..157D}). During FMG operation, one or a few limited positions in the range of the spectral line are fixed to observe Stokes parameters.

The magnetic-field calibration is a necessary and important process for producing reliable magnetic fields, and magnetic-field calibration methods vary for different types of instruments.
A Stokes polarimeter has ability to measure the full spectra of Stokes parameters $I$, $Q$, $U$, $V$ of a spectral line, then 
an inversion method is applied to derive the magnetic field under reasonable modes with other necessary thermal parameters. Since the full information of a spectral line is used by the Stokes polarimeter, it can measure the magnetic fields with higher accuracy, but lower temporal resolution than magnetograhp type instruments (\citeauthor{2008SoPh..249..167T}~\citeyear{2008SoPh..249..167T}; \citeauthor{2008SoPh..249..233I}~\citeyear{2008SoPh..249..233I}).
FMG obtain magnetograms at higher temporal resolutions than a spectragraph-base magnetograph, hence magnetic fields with rapid evolution can be studied by FMG observations.
For FMG, very limited Stokes information (one spectral position is fixed in routine observation) can be obtained, so the calibration of the magnetic field needs to be better and more carefully implemented.
The magnetic-field calibration method applied by FMG is linear calibration under the weak-field approximation (\citeauthor{1989ApJ...343..920J}~\citeyear{1989ApJ...343..920J}; \citeauthor{1991ApJ...372..694J}~\citeyear{1991ApJ...372..694J}). 
\begin{equation}
B_{\mathrm{L}}=C_{\mathrm{L}}V ,\\
\end{equation}
\begin{equation}
B_{\mathrm{T}}=C_{\mathrm{T}}(Q^{2}+U^{2})^{1/4}, \\
\end{equation}
\begin{equation}
\theta={\mathrm{arctan}}(B_{\mathrm{L}}/B_{\bot}),\\
\end{equation}
\begin{equation}
\phi=1/2{\mathrm{arctan}}(U/Q),\\
\end{equation}
where $B_{\mathrm{L}}$ and $B_{\mathrm{T}}$ are the line-of-sight (LOS) and transverse photosphere magnetic field, respectively.
$\theta$ is the inclination between the vector magnetic field and the direction normal to the solar surface and $\phi$ is the magnetic-field azimuth.
$C_{\mathrm{L}}$ and $C_{\mathrm{T}}$ are the calibration coefficients for the LOS and transverse magnetic fields, respectively.

The calibration coefficients $C_{\mathrm{L}}$ and $C_{\mathrm{T}}$ are decisive parameters for the determination of photosphere magnetic field, besides $C_{\mathrm{L}}$ and $C_{\mathrm{T}}$ are spectral-position related parameters (\citeauthor{1956PASJ....8..108U} \citeyear{1956PASJ....8..108U}; \citeauthor{1967IzKry..37...56R} \citeyear{1967IzKry..37...56R}; \citeauthor{1982SoPh...78..355L} \citeyear{1982SoPh...78..355L}). The ASO-S spacecraft run in a Sun-synchronous orbit with an altitude of about 720 km, an orbital inclination angle of 98.27°, and orbital period of about 99.2 mintes, which result in Sun--FMG radial velocity changes -3.9 to +3.9 km\,s$^{-1}$. The maximum value of the radial velocity corresponds to a wavelength shift from -0.07\AA~ to +0.07\AA~, which results in the changes of the observed spectral position.
As a result, calibration coefficients $C_{\mathrm{L}}$ and $C_{T}$ should be corrected at every moment in time based on the velocity of the ASO-S spacecraft.
Hence, to obtain and apply the calibration coefficients $C_{\mathrm{L}}$ and $C_{\mathrm{T}}$ depending on the velocity of ASO-S spacecraft would improve the FMG magnetic-field observation.

In this article, the longitudinal magnetic field of FMG and HMI
are investigated and used to calculate the calibration coefficients that related to spacecraft orbit velocity, with the purpose of improving the accuracy and reliability of the FMG magnetic-field measurement.
We present the observations in Section~\ref{sec:obser}; the methods and results analysis are given in Section~\ref{sec:method and results}; the discussions and conclusions are included in Section~\ref{sec:disc}.

\section{Observations}\label{sec:obser}
The FMG consists of the main elements of one birefringent Lyot filter (FWHM = 0.1 \AA, wavelength stabilization $\textless$ 0.02\AA) and two LCVR polarimeters 
to measure the magnetic field using the spectral line Fe {\sc i} $\lambda$5234.19 \AA.
The aperture of FMG is 14 cm and its effective focus length is 5355.76 mm. The spatial resolution of magnetic field observed by FMG is about 1.5 arcsec, for full-disk images 4k$\times$4k the spatial sampling is about 0.55 arcsec pixel. 
During routine observation its temporal resolution about two minutes and longitudinal sensitivity 15 G. Stokes $Q$, $U$, $V$ are observed by various combinations of azimuth and retardation of two LCVRs (cf. Table 1, \citeauthor{2020ChPhB..29l4211H} \citeyear{2020ChPhB..29l4211H}; \citeauthor{2019RAA....19..157D} \citeyear{2019RAA....19..157D}). Data processing includes dark field, flat field, polarization bias, and geometry
correction, then the polarization signals are calibrated to magnetic field under the weak-field approximation (\citeauthor{2019RAA....19..161S} \citeyear{2019RAA....19..161S}; \citeauthor{2021RAA....21...67X} \citeyear{2021RAA....21...67X}; \citeauthor{2022SoPh..297....6L} \citeyear{2022SoPh..297....6L}). The active-region data are cut out from full-disk images based on external auxiliary time and position information obtained from NOAA (\urlurl{www.solarmonitor.org/data/2023/04/01/meta/}).

\begin{figure}
\centerline{\includegraphics[width=1\textwidth,clip=]{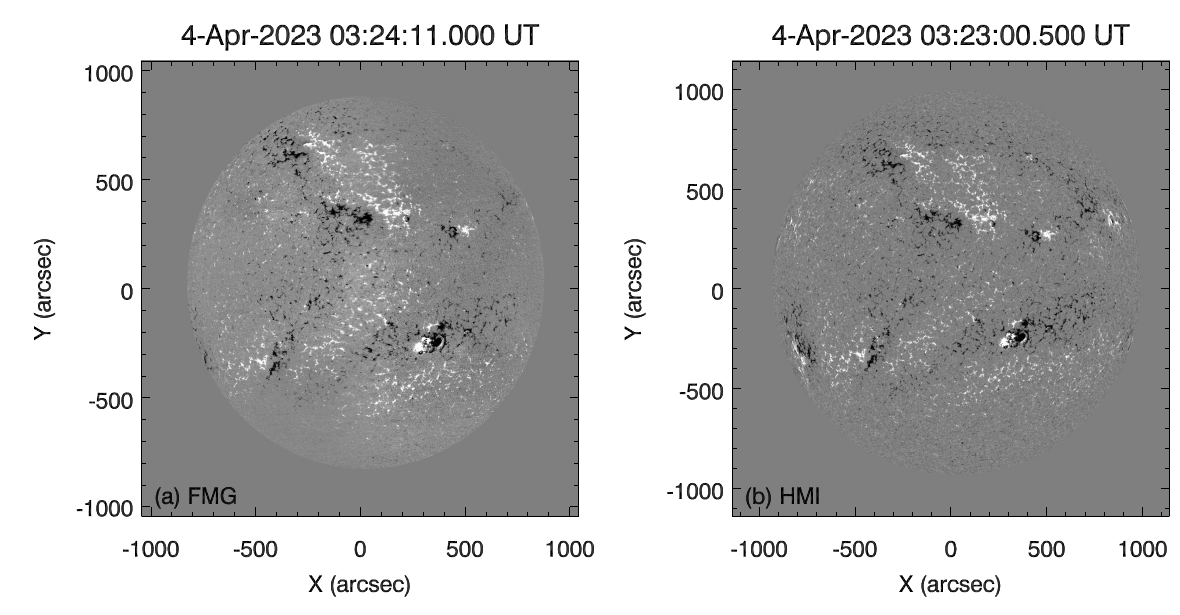}}
\caption{The original FMG ($left$) and HMI ($right$) full-disk longitudinal magnetic field observed 04 April 2023 at 04:24 UT quasi-simultaneously.} \label{fvfmghmiori}
\end{figure}

Helioseismic and Magnetic Imager is one of the instruments onboard Solar Dynamic Observatory (SDO/HMI: \citeauthor{2012SoPh..275..207S}, \citeyear{2012SoPh..275..207S}; \citeauthor{2012SoPh..275..229S}, \citeyear{2012SoPh..275..229S}).
HMI with a aperture of 14 cm and a 4k $\times$ 4k CCD observe full-disk magnetograms in the photospheric absorption line Fe {\sc i} centered at 6173.3~\AA{}. The spatial resolution of HMI is about 0.9 arcsec while the pixel size is about 0.5 arcsec. The temporal resolutions is 45 seconds for the longitudinal magnetic field, while for other observations there are different temporal resolutions (\urlurl{jsoc.stanford.edu/ajax/lookdata.html}). Stokes $Q$, $U$, $V$ are observed at six fixed wavelength positions by HMI, then inversion algorithms are applied to Stokes observations to deduce the photosphere magnetic field. In this article, the longitudinal magnetic field of FMG and HMI with temporal resolution of 45 seconds
are employed to investigate the calibration of FMG, with the aim to correct the coefficient of FMG modulated by and related to the spacecraft velocity. Figure \ref{fvfmghmiori} shows an example of an original FMG and HMI longitudinal magnetic field observe at 04 April 2023 03:24 UT quasi-simultaneously.
After geometric correction using position information during the observation, these two longitudinal magnetic-field signals have very good consistency, however there are still differences in the size and spatial resolution for these two images.

\section{Methods and Results}\label{sec:method and results}
\begin{figure}
\centerline{\includegraphics[width=1\textwidth,clip=]{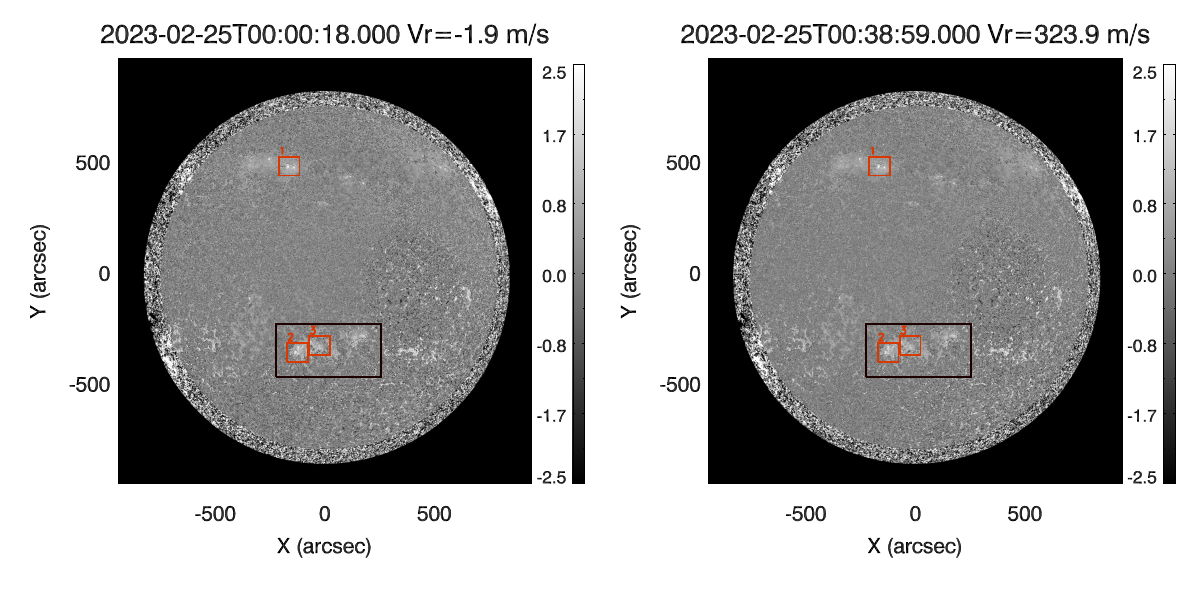}}
\caption{The distribution of the ratios of line-of-sight magnetic field between the FGM and HMI for $v_r$ = -1.9 m\,s$^{-1}$ ($left$) and $v_r$  = 323.9 m\,s$^{-1}$ ($right$) labeled, respectively.
The region indicated by a $black$ $rectangle$ is the sub-region shown in Figure \ref{fd_cross_sub}. The sub-regions 1, 2, and 3 indicated by a $orange$ $rectangle$ are the active regions for calculating
pixel-to-pixel calibration coefficients in Figure \ref{plotsercoeffarrvsvrarrvsvr}. } \label{fd_cross}
\end{figure}

\begin{figure}
\centerline{\includegraphics[width=1\textwidth,clip=]{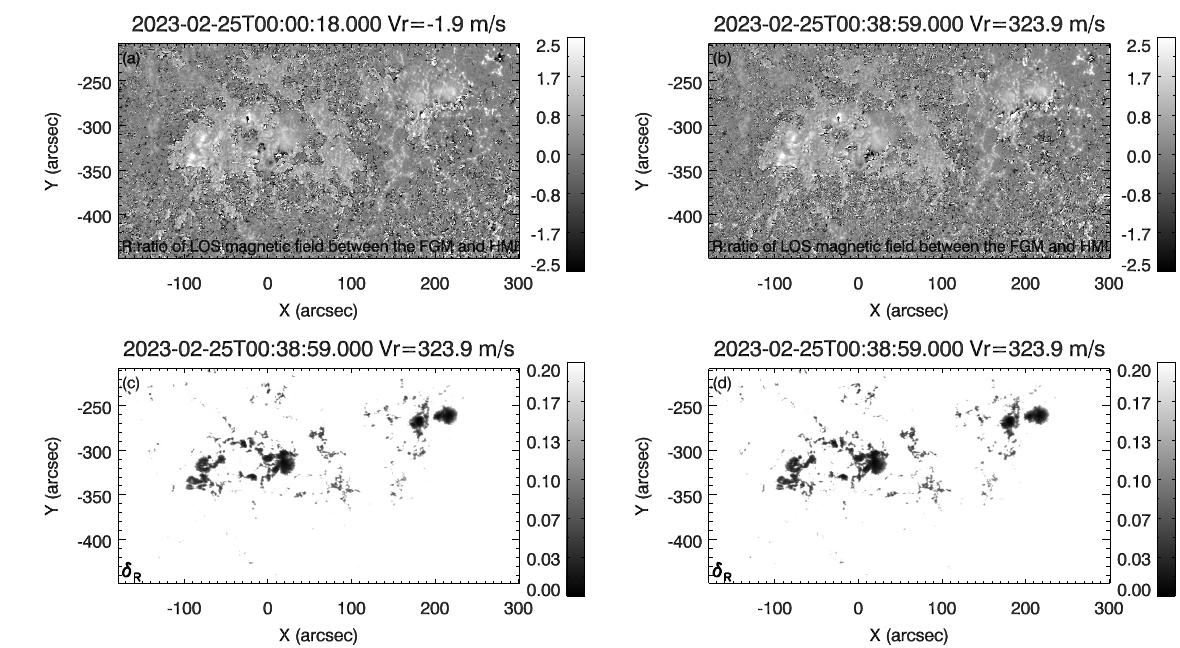}}
\caption{Panels \textbf{a} and \textbf{b} are the ratio of line-of-sight magnetic field between the FGM and HMI, which are same as Figure \ref{fd_cross}, but for a sub-region cut from full-disk image in Figure \ref{fd_cross} indicated by a $black$ $rectangle$. Panels \textbf{c} and \textbf{d} are the errors of the ratio of line-of-sight magnetic field between the FGM and HMI for Panels \textbf{a} and \textbf{b}, respectively.} \label{fd_cross_sub}
\end{figure}

\begin{figure}
\centerline{\includegraphics[width=1\textwidth,clip=]{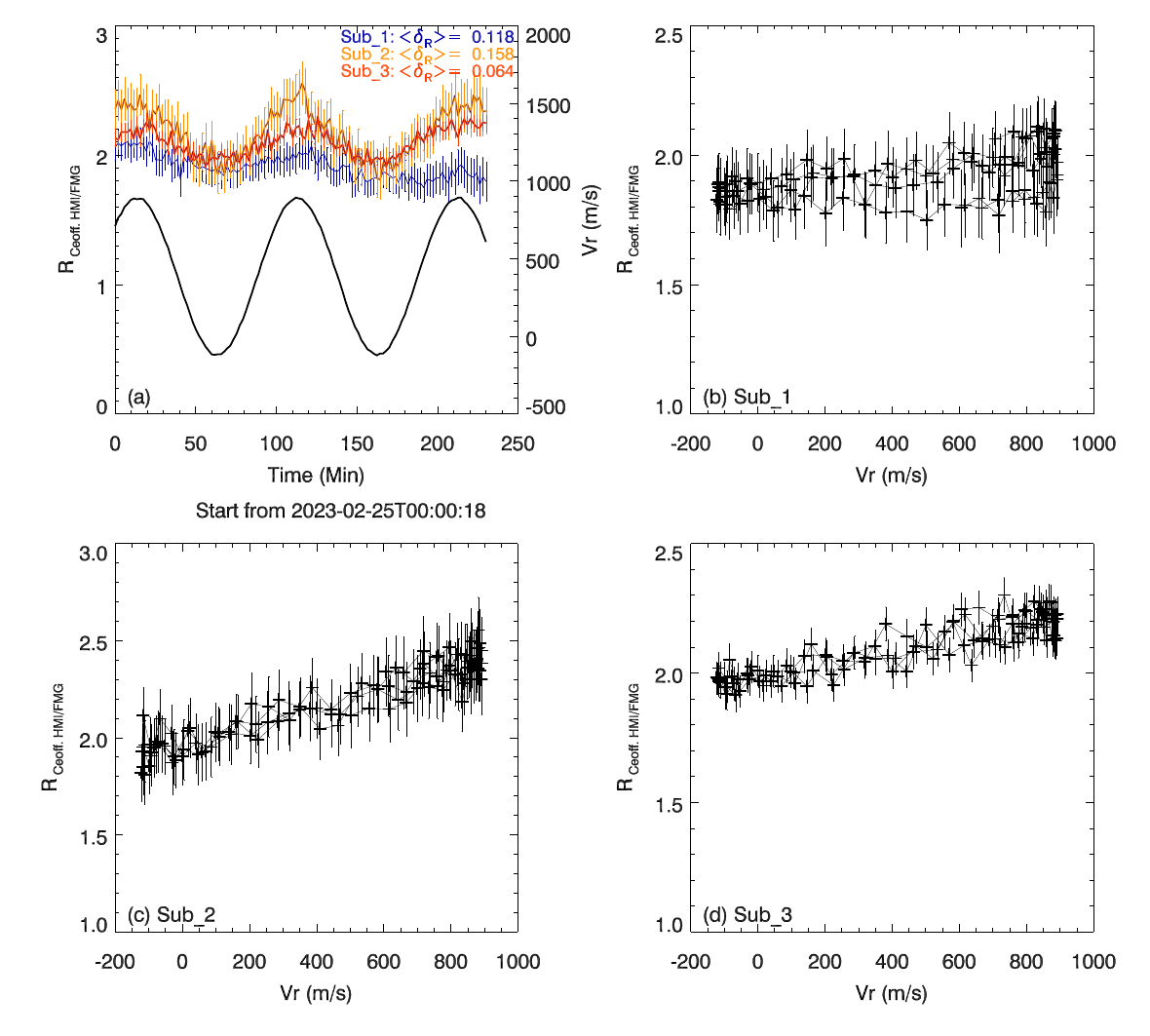}}
\caption{\textbf{(a)} the time series of cross coefficents for sub-regions 1, 2, and 3 indicated in Figure \ref{fd_cross}, here $v{_r}$ with the unit of m\,s$^{-1}$ (line-of-sight component of spacecraft velocity) is plotted as an additional left $y$-axis. 
\textbf{(b)}, \textbf{(c)} and \textbf{(d)} are the scatter plots of cross coefficents vs. $v{_r}$ for sub-regions 1, 2, and 3 as in Figure \ref{fd_cross}, respectively.}\label{plotsercoeffarrvsvrarrvsvr}
\end{figure}

In order to compare calibration coefficients of FMG and HMI, firstly these two magnetic-field images are co-aligned by adjusting the size and pixel resolution.
Based on observational information, especially coordinate information, the two magnetic field images are basically aligned when they are converted to a heliospheric coordinate system.
However, due to the exist of various system errors, there may still be slight position deviations between the two magnetic-field images of FMG and HMI. Therefore, an image-alignment method of a scale invariant feature transform (SIFT: \citeauthor{790410} \citeyear{790410}) based on image features was adopted to co-align FMG and HMI magnetic-field images. 
The three main steps of SIFT for image matching include: i) the feature-point identification; ii) the feature-point matching; iii) determination of image-registration parameters. Feature-point identification
searching for rotation, scaling, and translation at different image spatial scales of invariant feature points. Hence, SIFT is suitable to solve the image rotation, affine
transformations, intensity, and viewpoint changes in matching features. The SIFT method has been successfully applied in solar image processing (\citeauthor{yp2018} \citeyear{yp2018}; \citeauthor{jkf2019} \citeyear{jkf2019}). After co-alignment of the FMG and HMI magentic field, the calibration coefficients of FMG corresponding to HMI can be obtained pixel-to-pixel on the solar surface directly.
Figure \ref{fd_cross} shows the full-disk distribution of the ratio of line-of-sight magnetic field between the FGM and HMI for line-of-sight velocity $v_r$ = -1.9 m\,s$^{-1}$ and $v_r$ = 323.9 m\,s$^{-1}$, respectively. While Panels a and b in Figure \ref{fd_cross_sub} show the enlarged images cut from Figure \ref{fd_cross} indicated by a black rectangle to investigate the distribution of coefficients of FMG comprising to HMI in detail, Panels c and d in Figure \ref{fd_cross_sub} give the errors of these ratios correspond Panels a and b, respectively. From Figure \ref{fd_cross} and \ref{fd_cross_sub}, it is can be found that the distribution of ratio of line-of-sight magnetic field between the FGM and HMI within the activity region is relatively flat and consistent, which can also be seen from the errors of these ratios shown in Panels c and d in Figure \ref{fd_cross_sub}, where the relative errors are small in active regions. The fluctuation of these ratios in the quiet region is relatively large, as some magnetic field with small value in the quiet region maybe close to the noise level.
Hence, the comparison of calibration coefficients between FMG and HMI is focused on active region, to obtain the reliable results. In Figure \ref{fd_cross}, three small sub-regions with strong magnetic field labeled 1, 2, and 3 are chosen to investigate the ratio of line-of-sight magnetic field between FGM and HMI. A series of full-disk images co-aligned by SIFT, the same as Figure \ref{fd_cross} are rotated to a reference time, and the sub-regions are cut out from full-disk images by their pixel positions, then the 
median values of ratio of line-of-sight
magnetic field between the FGM and HMI for these series of sub-regions, which can indicate the effective values of these ratios on the whole, are calculated. Figure \ref{plotsercoeffarrvsvrarrvsvr} a shows the ratio of line-of-sight magnetic field between the FGM and HMI of these time series for sub-regions 1, 2, and 3, and $v_r$ with the unit of m\,s$^{-1}$ (line-of-sight component of spacecraft velocity) is plotted as an additional left $y$-axis to investigate the relations between calibration coefficients and the spacecraft velocity. Here the mean of errors of these ratios for sub-regions are calculated and labeled $\delta_{R}$, with the values of $\delta_{R}$=0.118, 0.158, and 0.064 for sub-regions of 1, 2, and 3, respectively. Figure \ref{plotsercoeffarrvsvrarrvsvr} b, c, and d show the scatter plots of ratio of line-of-sight
magnetic field between the FGM and HMI vs. $v_r$ for sub-regions 1, 2, and 3 labeled. It can be found that the ratios of line-of-sight
magnetic field between the FGM and HMI have a consistent trend with the longitudinal velocity of the spacecraft, and the monotonicity of the ratio and spacecraft longitudinal velocity can also be clearly seen from b, c, and d. It means that the spacecraft longitudinal velocity regularly modulates the magnetic-field calibration coefficient due to the shift of position of the observed spectral line by Doppler effect resulting from spacecraft velocity.

\begin{figure}
\centerline{\includegraphics[width=1\textwidth,clip=]{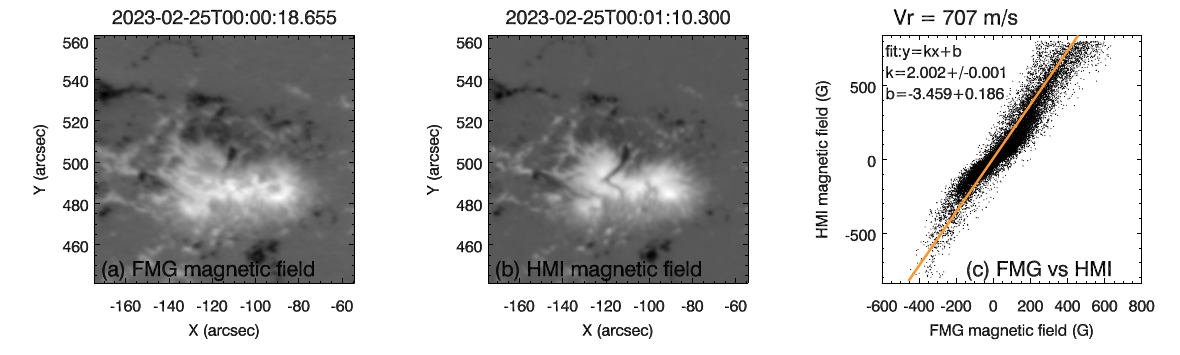}}
\caption{ \textbf{(a)} and  \textbf{(b)} active-region longitudinal magnetic field of FMG and HMI.  \textbf{(c)} the scatter plot of FMG Panel \textbf{a} at spacecraft velocity of $v_r$ of 707 m\,s$^{-1}$ and HMI Panel \textbf{b} active-region magnetic field, the $yellow$ $line$ is the fit $y=kx+b$
.} \label{subregionfit}
\end{figure}

The reliability of the ratio of line-of-sight magnetic field by pixel-to-pixel comparison of FMG and HMI in Figures \ref{fd_cross} and \ref{fd_cross_sub} is severely affected by signal noise. 
An alternative method is fit the active region magnetic field as a whole to obtain the ratios between FMG and HMI.
Firstly, the active region or sub-region with strong magnetic field of FMG and HMI are plotted as scatter plots; then linear fitting ($y=kx+b$) is applied to these scatter plots to calculate the slope of this linear formula, the slope fitted can be regarded as ratio of the coefficient between FMG and HMI. Figure \ref{subregionfit} shows the process to calculate the whole fitting ratio between FMG and HMI, Panels a and b are the longitudinal magnetic field of FMG and HMI observed quasi-simultaneously. In Panel c the scatter plot of FMG at spacecraft velocity of $v_r$ of 707 m\,s$^{-1}$ and HMI magnetic field are plotted, the yellow line is the fitting result, here the slope $k$ with value of 2.002 $\pm$ 0.001 is regarded as the ratios of coefficient for this sub-region on the whole when $v_r$ is 707 m\,s$^{-1}$. For this fitting method, the ratio of coefficients is fitted through overall distributions of the scatter plot, additionally the erorrs both of FMG and HMI are included in the fitting process. Hence the effects of signal noise and size of region are avoided, which is the advantage of this fitting method. 

\begin{figure}
\centerline{\includegraphics[width=1\textwidth,clip=]{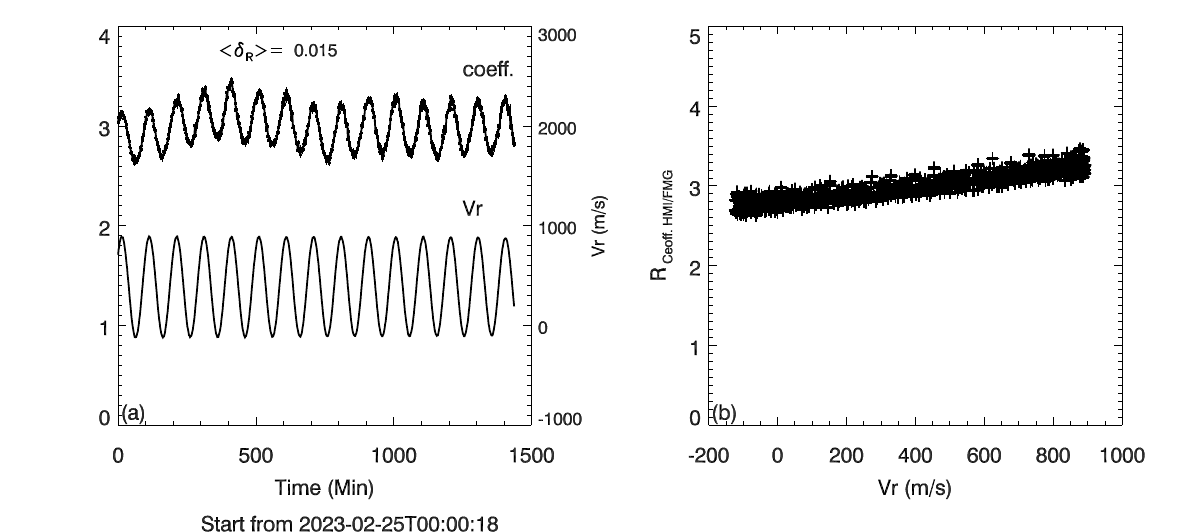}}
\caption{\textbf{(a)} the time series of cross coefficient of HMI/FMG using fitting method and ${v_r}$ as an additional left $y$-axis on 25 February 2023 for a sub-region with strong magnetic field.
\textbf{(b)} are the scatter plots of cross coefficents vs. $v_r$.}\label{plotsercoeffarrvsvrarrvsvrfitmethod}
\end{figure}

Figure \ref{plotsercoeffarrvsvrarrvsvrfitmethod} shows a time series of the ratios of the coefficient between FMG and HMI using fitting method and $v_r$ during a period of 25 February 2023 for a sub-region with strong magnetic field. 
It can be found that the ratios of coefficient are modulated by the longitudinal velocity of the spacecraft [$v_r$], with evident periodicity that is the same exactly as orbital period.
The scatter plot with a high concentration of points means that the one-to-one correspondence between calibration coefficient and spacecraft velocity is excellent 
during this observation with multiple spacecraft orbital period lasting about one day.

\begin{figure}
\centerline{\includegraphics[width=1\textwidth,clip=]{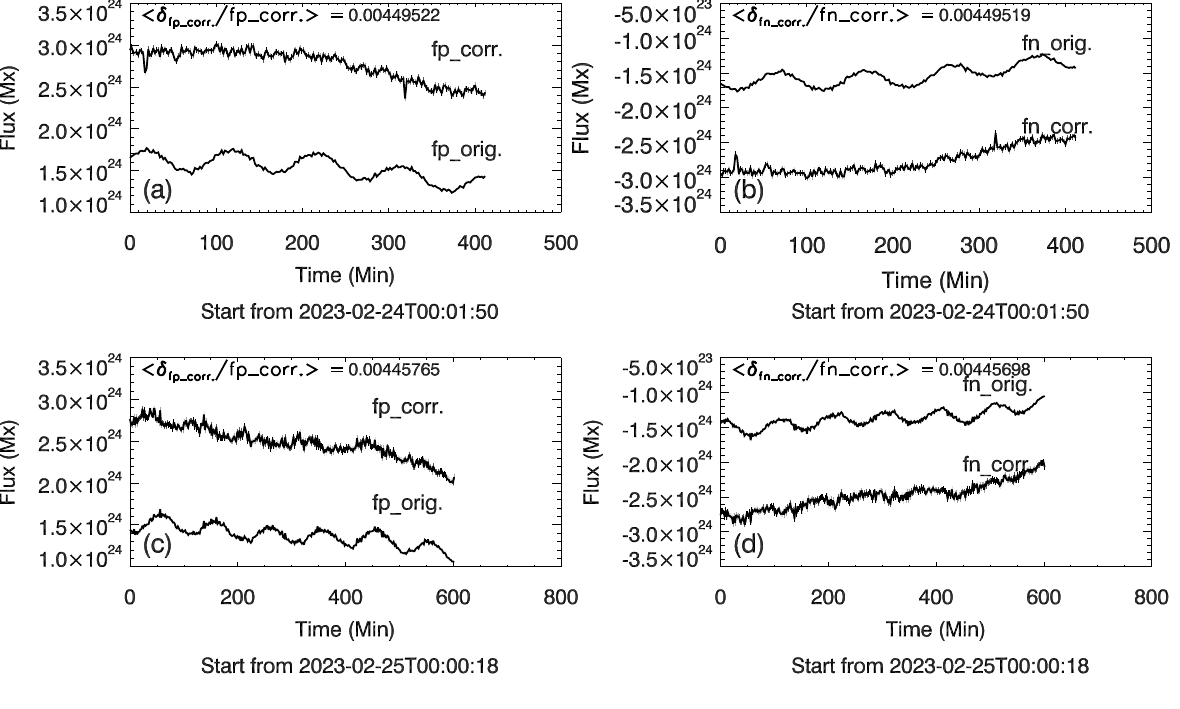}}
\caption{The examples of the evolution of the magnetic flux for active regions before and after cross-calibration coefficients correction. 
Here, the cross-calibration coefficients are calculated by fitting. Panels \textbf{a} and \textbf{c} are active region NOAA 13230 observed on 24 February 2023, Panels \textbf{a} and \textbf{c} are NOAA 13236 observed on 25 February 2023, positive and negative magnetic flux before and after correction are indicated by $fp_\mathrm{orig}$, $fp_\mathrm{corr}$, $fn_\mathrm{orig}$, and $fn_\mathrm{corr}$, respectively. For plots of $fp_\mathrm{corr}$ and $fn_\mathrm{corr}$, the error bars are added.} \label{fluxcorrect}
\end{figure}

To correct FMG magnetic-field values modulated/affected by spacecraft velocity is the ultimate goal of obtaining ratios of calibration coefficients between FMG and HMI.
Figure \ref{fluxcorrect} shows some examples of the evolution of magnetic flux for active regions before and after correcting by fitting method, Panels a and b are active region NOAA 13230 observed on 24 February 2023, Panels c and d are NOAA 13236 observed on 25 February 2023, 
the evolution of positive and negative magnetic flux before and after correction are plotted and labeled by $fp_\mathrm{orig}$, $fp_\mathrm{corr}$, $fn_\mathrm{orig}$, and $fn_\mathrm{corr}$, respectively.
Additionally, the error bars are calculated and plotted for the corrected magnetic flux $fp_\mathrm{corr}$ and $fn_\mathrm{corr}$ in each panel, here the errors of $fp_\mathrm{corr}$ and $fn_\mathrm{corr}$ are about 0.45\% comparing the values of $fp_\mathrm{corr}$ and $fn_\mathrm{corr}$ .
From Figure \ref{fluxcorrect}, it can be found that the evolution of both positive and negative magnetic flux for these active region before correction are modulated by spacecraft velocity evidently with the same periodicity. 
While after the correction, the periodicities of evolutions of the magnetic flux become weaker significantly.
Therefore, the coefficients corrected FMG magnetic field will be more suitable for quantitative research on solar magnetic field.

\section{Discussions and Conclusions}\label{sec:disc}
The FMG onboard the ASO-S spacecraft operates in Sun-synchronous orbit with orbital period about 99.2 minutes. Subsequently Sun--FMG radial velocities also have periodicity with longitudinal velocity [$v_r$] between Sun and FMG changes periodically with from -3.9 to +3.9 km\,s$^{-1}$.
Longitudinal velocity [$v_r$] of FMG resulting Doppler velocity affects the calibration of magnetic field obviously for routine FMG observations, when one spectral position is fixed during an observation.
The method of correcting the magnetic field affected by spacecraft velocity can be investigated by comparing the calibration coefficients with other instruments. 

In this article, HMI longitudinal magnetic field with 45-second cadence working as reference to calibrate FMG magnetic field is tested.
From the test results in Figure \ref{fluxcorrect}, it can be found that this type of calibration method can largely eliminate the influence of spacecraft periodic velocity on magnetic field of FMG.
The corrected magnetic field is closer to the actual conditions, which would definitely improve the reliability of the FMG magnetic field.
The consistency with HMI magnetic field is not the goal of FMG calibration, the real goal is to eliminate the impact of spacecraft orbital velocity period on the FMG magnetic field, and to have a precise measurement of magnetic field without biases or systematic errors.
For example Figure \ref{fluxcorrect}, the effect of magnetic flux evolution being consistent with spacecraft velocity period should be corrected, as it is a false signal. Since the linear calibration method is adopted by FMG, and the consistency of FMG magnetic-field values itself without the influence of spacecraft orbital period is very important.

FMG observes magnetic field over the full solar disk, and calibration coefficient is very sensitive to the position on the solar surface.
There exists a large-scale background Doppler velocity field related to the solar position, which inevitably participates in the influence of spacecraft velocity on the FMG magnetic field.
Therefore, a complete data sample, in which the active-region observations exist positions on the solar surface, is a necessity in order to fully and effectively correct the FMG magnetic field influenced by the periodicity of spacecraft velocity. However in this article, the focus of attention has been placed on the impact of spacecraft velocity.

Another aspect that needs to be mentioned is that calibration coefficients are spectral position related, since the relation between Stokes signal strength and magnetic field varies for different observation position of a spectral line when periodic velocity affects the shape of spectral lines.
In the routine observations of FMG at present, the observed spectral line is fixed at -0.08 \AA~away from the center of the line. 
So the dateset of correcting calibration coefficients in this article are suitable to correct the FMG magnetic that observed at line wing of -0.08 \AA.
Additionally, dateset of calibration coefficients for the observation at line wing of -0.08 \AA~ is not complete, and more active regions try to cover solar disk are need.

An alternative method to correct FMG magnetic field is a post-processing method based on the periodicity of spacecraft velocity with high certainty. For example evolution of magnetic flux in Figure \ref{fluxcorrect},
filtering technology (such as the FFT) can be applied to these curves to obtain the component with the period of 99.2 minutes, then a reliable magnetic field can be deduced by eliminating 99.2 minutes periodic component.
Additionally, machine-learning methods of artificial intelligence may also correct the FMG magnetic field after the acquisition of a large observation sample with enough information of observation spectral position, solar surface position, and a full range of spacecraft velocity.

From the results in the article, it can be found that through cross comparison with HMI, it is largely possible to correct the periodic influence of spacecraft orbital velocity on the FMG magnetic field, and obtain a more realistic magnetic field observed by FMG. However, it should be noted that HMI is not the only sample used as a reference standard, since the self consistency of the FMG magnetic field avoiding spacecraft velocity impact is the main goal of correction. Another observation can be chosen as reference is data obtained by SMAT ($Solar$ $Magnetism$ and $Activity$ $Telescope$ at the Huairou Solar Observing Station), the advantage of FMG and SMAT is that they have common design parameters. To correct FMG magnetic field more comprehensively, the main aspect is to establish a more comprehensive calibration database with adequate information of observation spectral position, solar-surface position and full range of spacecraft velocity, all of these need a continuous long-term stable observation of FMG.

\acknowledgments
This work is supported by National Key R\&D Program of China 2022YFF0503000 (2022YFF0503001),the Strategic Priority Research Program
on Space Science, the Chinese Academy of Sciences (Grant No. XDA15320301, XDA15320302,
XDA15052200), Natural Science Foundation of China (Grant No 11203036,11703042, U1731241,11427901, 11473039, 11427901 and 11178016), the
Young Researcher Grant of National Astronomical Observations,
Chinese Academy of Sciences, and the Key Laboratory of Solar
Activity National Astronomical Observations, Chinese Academy

\end{document}